\documentclass{elsart}
\usepackage{natbib}
\usepackage{epsfig}
\usepackage{amssymb}

\begin{document}

\begin{frontmatter}

\title{On Estimation of Hurst Scaling Exponent through Discrete Wavelets}

\author{P. Manimaran}
\address{School of Physics, University of Hyderabad,
 Hyderabad 500 046, India}
\author{Prasanta K. Panigrahi$^{1,2}$, and}
\ead{prasanta@prl.res.in}
\author{Jitendra C. Parikh$^2$}
\address{$^1$Indian Institute of Science Education and Research (Kolkata), Salt Lake,
Kolkata 700 106, India}
\address{$^2$Physical Research Laboratory, Navrangpura, Ahmedabad
380 009, India}

\begin{abstract}
We study the scaling behavior of the fluctuations, as extracted
through wavelet coefficients based on discrete wavelets. The
analysis is carried out on a variety of physical data sets, as well
as Gaussian white noise and binomial multi-fractal model time series
and the results are compared with continuous wavelet based average
wavelet coefficient method. It is found that high-pass coefficients
of wavelets, belonging to the Daubechies family are quite good in
estimating the true power in the fluctuations in a non-stationary
time series. Hence, the fluctuation functions based on discrete
wavelet coefficients find the Hurst scaling exponents accurately.

\end{abstract}


\end{frontmatter}

\section{Introduction}
Fractals exhibiting self-similar behavior, are ubiquitous in nature \cite{mandel}. They manifest in
areas ranging from financial markets to natural sciences. Several
techniques have been developed to study the correlations and
scaling properties of time series exhibiting self-similar
behavior; some of these data sets are non-stationary in character.
Various methods like R/S analysis \citep{hur}, structure function
method \cite{feder}, wavelet transform modulus maxima \cite{arn1},
detrended fluctuation analysis and its variants
\cite{khu,gopi,ple,chen,matia,krs,phand,xu,peng,net}, average
wavelet coefficient method \cite{awc} and a recently developed
discrete wavelet based approach by the present authors, have been
employed for the characterization of fluctuations
\cite{mani1,mani2}.

Wavelets, through their multi-resolution and localization abilities,
are well suited for extracting fluctuations at various scales from
local trends over appropriate window sizes \cite{daub,mall}. The
nature of the fluctuations extracted partly depend on the choice of
the wavelets, which are designed to have properties useful for a
desired analysis. For example, the Daubechies family of wavelets
satisfy vanishing moment conditions, which make them ideal to
separate polynomial trends in a data set. In a number of wavelet
based approach for characterizing self-similar data, wavelet
high-pass coefficients are used in finding the fluctuation function
or other related quantities. For example, in the average wavelet
coefficient method, continuous wavelet transform is used to find the
variations through the corresponding wavelet coefficients, from
which the Hurst scaling exponent is computed.

The goal of the present article is to analyze the self-similar
properties of the fluctuations, as extracted through discrete
wavelet coefficients, for the purpose of checking its efficacy
vis-a-vis the continuous wavelet based method. The study is carried
out on Gaussian white noise and binomial multifractal model time
series, as well as a number of physical data sets. It is found that
the fluctuation function based on discrete wavelet coefficients is
accurate in estimating Hurst scaling exponent. It is worth
emphasizing that continuous wavelets, because of their over
completeness tend to overestimate the power. In comparison, the
discrete wavelets provide a complete orthonormal basis, ensuring
that fluctuations are independent at each level.

The paper is organized as follows. In the following section, we
study the nature of the fluctuations extracted through the wavelet
coefficients. In Sec.III, we then proceed to the detailed analysis
of fluctuations of Gaussian white noise and binomial multi-fractal
model and compare the discrete wavelet based result with average
wavelet coefficient method. Subsequently, we carry out analysis of
the fluctuations in the data of observed ionization current and
potentials in tokamak plasma, time series constructed from random
matrix ensembles and the financial data sets belonging to NASDAQ and
Bombay stock exchange (BSE) composite indices. Finally, we conclude
after summarizing our findings and giving future directions of work.

\section{Fluctuations in the wavelet domain}
In discrete wavelet transform it is well-known that, a given
signal belonging to $L^2$ space can be represented in a nested
vector space spanned by the scaling functions alone. This basic
requirement of multi-resolution analysis (MRA) can be formally
written as \cite{ram},
\begin{equation}
...\subset \nu_{-2} \subset \nu_{-1}\subset \nu_{-0}\subset
\nu_{1}\subset \nu_{2}...\subset L^2,
\end{equation}
with $\nu_{-\infty} = {0}$ and $\nu_{\infty}= L^2$. This provides a
successive approximation of a given signal in terms of low-pass or
approximation coefficients. It is clear that, the space that
contains high resolution signals will also contain signals of lower
resolution. The signal or time series can be approximated at a level
of ones choice, for use in finding the local trend over a desired
window. The fluctuations can then be obtained by subtracting the
above trend from the signal. We have followed this approach earlier
for extracting the fluctuations, by elimination of local polynomial
trends through the Daubechies wavelets \cite{mani1,mani2}.

 Wavelets also provide a decomposition of a signal in terms of wavelet
coefficients and one low-pass coefficient:

\begin{equation}
L^2 = ~~...\oplus \textit{W}_{-2} \oplus \textit{W}_{-1} \oplus
\textit{W}_{0} \oplus \textit{W}_{1} \oplus \textit{W}_{2}~~...
\end{equation}
and
\begin{equation}
\textit{W}_{-\infty} ~~ \oplus ~~...~~\oplus~~\textit{W}_{-1} =
\nu_0.
\end{equation}

Wavelet coefficients represent variations of the signal at different
scales. For example, level one coefficients capture the highest
frequency components, corresponding to variations at highest
resolution and other wavelet coefficients represent variations at
progressively higher scales or lower resolutions. As mentioned
earlier, these coefficients can differ significantly from the true
fluctuations in the data sets. Below, we explore this aspect through
the estimation of Hurst exponents.

Let $x_t$ ($t=1,2,3,...,N$) be the time series of length $N$.
First one determines the "profile" (say $Y(i)$), which is
cumulative sum of series after subtracting the mean.
\begin{equation}
Y(i) = \sum_{t=1}^i [x_t - \langle x \rangle], ~~~ i=1,....,N.
\end{equation}

Next, we obtain the statistics of scale dependence by transforming
the profile of the time series into wavelet space, the
coefficients of wavelets at various scales $s$ are used to
determine the fluctuation function. The high frequency details are
captured by the lower scale wavelet coefficients and the higher
scales capture the low frequency details. By convolving the
discrete wavelet transform over the given time series, the wavelet
coefficients are obtained for various scales:

\begin{equation}
W_{j,k} = 2^{j/2} \sum_{i=0}^{N-1} Y_i ~~\psi(2^j t - k).
\end{equation}

Here '$j$' is the scaling index and $k$ represents the translation
variable. Since discrete wavelet transform satisfies orthogonality
condition, it can provide the information of time series at
various scales unambiguously. Performing wavelet transform using
Daubechies basis, the polynomial trends in the time series are
eliminated. In the analysis carried out below we make use of the
Daubechies-4 wavelets. As has been observed earlier, small
fluctuations are least affected by this basis and straight line
trends (akin to a linear fit) are removed through the use of this
wavelet \cite{mani1}.

The wavelet power is calculated by summing the squares of the
coefficient values for each level:

\begin{equation}
A(j)= \sum_{k=0}^{\frac{N}{2^j}-1} W_{j,k}^2.
\end{equation}

To characterize the time series, the fluctuation function $F(s)$
at a level $s$ is obtained from the cumulative power spectrum:

\begin{equation}
F(s) = \left [ \sum_{j=1}^{s} A(j) \right ]^{1/2}.
\end{equation}

The scaling behavior is then obtained through,

\begin{equation}
F(s) \sim s^H.
\end{equation}

Here $H$ is the Hurst scaling exponent, which can be obtained from
the slope of the log-log plot of $F(s)$ vs scales $s$. It is well
known that Hurst exponent is one of the fractal measures, which
varies from $0 < H < 1$. For persistent time series $H > 0.5$ and
$H=0.5$ uncorrelated series. $H < 0.5$ for anti-persistent time
series.
\begin{figure}
\centering
\includegraphics[width=3in]{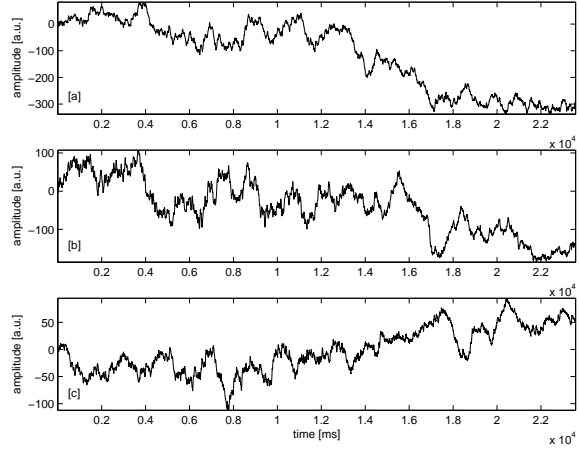}
\caption{Time series of (a) ion saturation current (IC), (b)
floating potential (FP), $6~mm$ inside the main plasma and (c) ion
saturation current (ISC), when the probe is in the limiter shadow.
Each time series is of approx. $24,000$ data points.}
\end{figure}

\begin{figure}
\centering
\includegraphics[width=3in]{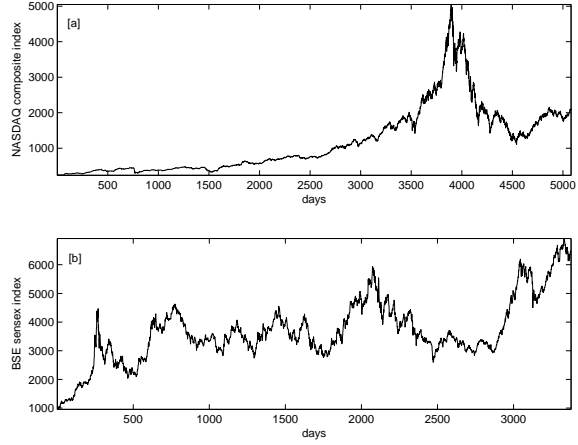}
\caption{ Time series of (a) NASDAQ composite index for a period of
20 years, starting from 11-Oct-1984 to 24-Nov-2004, and (b) BSE
sensex index, over a period of 15 years, starting from 2-Jan-1991 to
12-May-2005.}
\end{figure}

\begin{figure}
\centering
\includegraphics[width=3in]{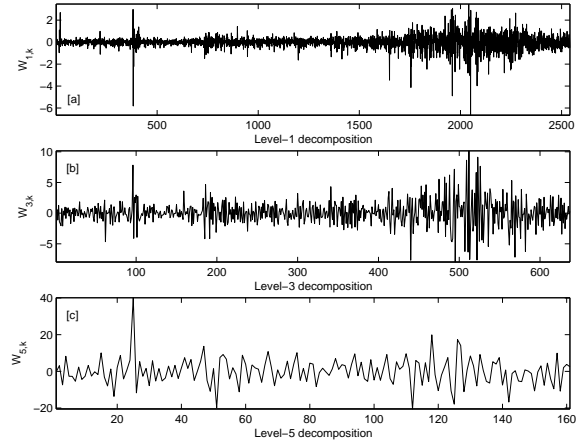}
\caption{Wavelet coefficients at various levels for NASDAQ
composite data.}
\end{figure}

\begin{figure}
\centering
\includegraphics[width=3in]{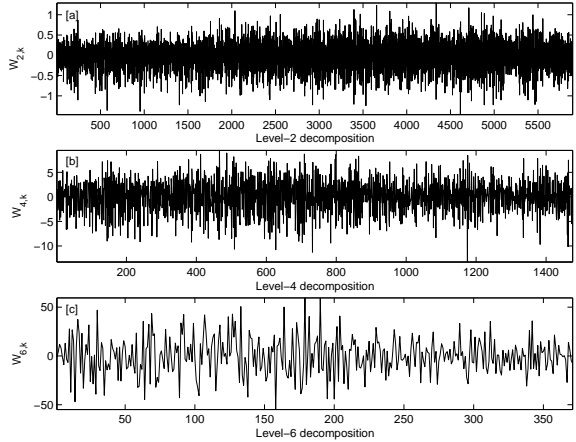}
\caption{Wavelet coefficients at various levels for tokamak plasma
data involving ion saturation current from top to bottom
respectively.}
\end{figure}

We have analyzed three sets of experimentally observed time series
of variables in ohmically heated edge plasma in Aditya tokamak
\cite{jos}. The time series are i) ion saturation current, ii) ion
saturation current when the probe is in the limiter shadow, and iii)
floating potential, 6mm inside the main plasma. Each time series has
about 24,000 data points sampled at 1MHZ \cite{jha}. These are shown
in Fig. 1. The study of fluctuations play an important role in our
understanding of turbulent transport of particles and heat in the
plasma. In Fig. 2, we show financial time series of NASDAQ composite
index and BSE sensex index values. Wavelet coefficients at various
scales have been displayed in Fig. 3 and Fig. 4. In Figs. 5 and 6,
we have shown $F(s)$ versus $s$ for the time series of three
experimental data sets and financial stock market data,
respectively. The scaling exponent $H$, for all the three
experimental time series as well as financial data sets shows long
range correlations ($H > 0.5$). We have also analyzed the discrete
time series obtained from random matrix ensembles corresponding to
Gaussian orthogonal ensemble (GOE), Gaussian symplectic ensemble
(GSE) and Gaussian unitary ensemble (GUE). These show long range
anti-correlation behaviors $H < 0.5$. Gaussian  diagonal ensemble
(GDE) shows uncorrelated behavior, $H = 0.5$. It is worth mentioning
that, we have followed the recent approach of Ref. \cite{rela}, for
converting the random matrix ensemble data to discrete time series
\cite{ran}. In Table-I and II, Hurst exponents of various data sets
are given. These results agree with our previous discrete wavelet
based approach.
\begin{figure}
\centering
\includegraphics[width=3in]{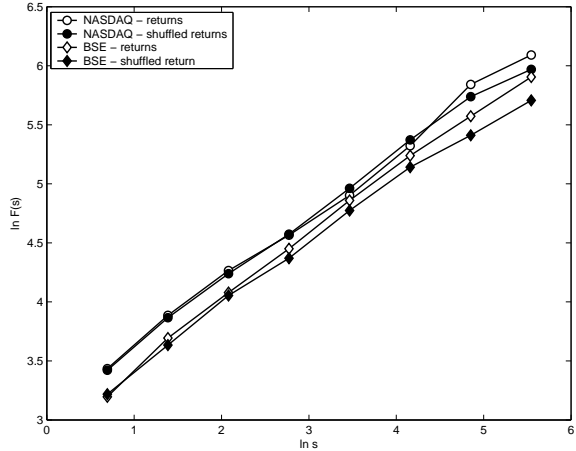}
\caption{The log-log plot of fluctuation function $F(s)$ vs $s$, for
the time series of NASDAQ composite index and BSE sensex index
values for returns and shuffled returns. One clearly sees long range
correlation behavior.}
\end{figure}

\begin{figure}
\centering
\includegraphics[width=3in]{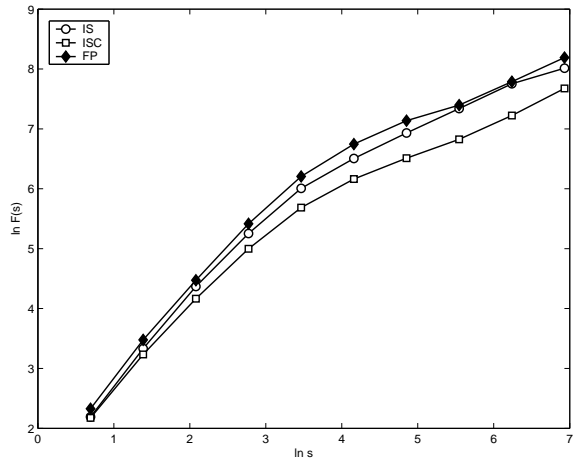}
\caption{The log-log plot of fluctuation function $F(s)$ vs $s$, for
the time series of tokamak plasma data. For larger window sizes one
observes long-range correlations.}
\end{figure}

\begin{figure}
\centering
\includegraphics[width=3.2in]{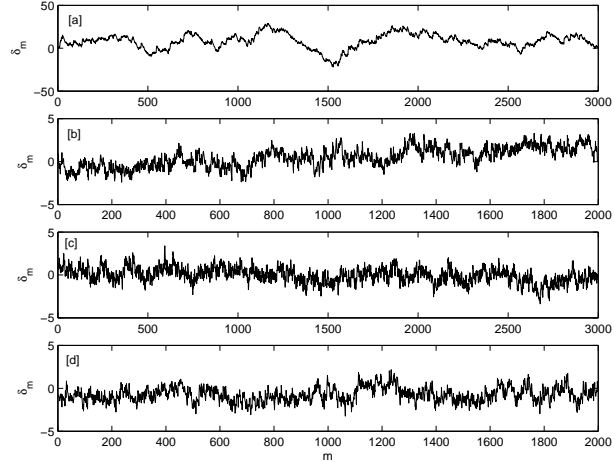}
\caption{Time series obtained from Gaussian random ensembles [a]
GDE, [b] GSE, [c] GOE, and [d] GUE. For GDE $H \sim 0.5$ and for
others one sees the long range anti-correlation behavior $H <
0.5$}
\end{figure}

\begin{figure}
\centering
\includegraphics[width=3in]{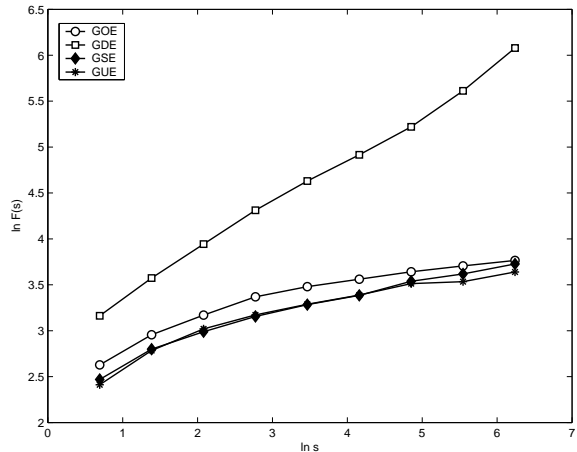}
\caption{The log-log plot of fluctuation function $F(s)$ vs $s$,
for the time series involving Gaussian random ensembles.}
\end{figure}

\begin{table}
\centering
\begin{tabular}{|c|c|}
\hline
Data & Hurst (H) \\
\hline
NASDAQ - returns          & 0.553 \\
NASDAQ - shuffled returns & 0.542 \\
BSE - returns             & 0.548 \\
BSE - shuffled returns    & 0.518 \\
\hline
IC  & 0.585 \\
ISC & 0.554 \\
FP  & 0.549 \\
\hline
GOE& 0.095 \\
GDE& 0.495 \\
GSE& 0.143 \\
GUE& 0.107 \\
\hline
\end{tabular}
\caption{Computed Hurst exponent for various data sets, involving
financial, tokamak plasma and random matrix energy fluctuations.}
\end{table}

\begin{table}
\centering
\begin{tabular}{|c|c|c|c|c|}
\hline \hline
  Data & $H_{a}$ & $H_{d}$ & $H_{c}$ \\
  \hline \hline
  BMF  & 0.8390    &0.8421     &0.8433\\
  WGN  & 0.5       &0.5077     &0.5163\\
  \hline \hline
\end{tabular}
\caption{Hurst scaling exponent computed analytically ($H_a$),
through discrete wavelet coefficient method ($H_d$)and continuous
average wavelet coefficient method ($H_c$) for both White Gaussian
Noise(WGN) and Binomial Multi-Fractal Model (BMF) time series.}
\end{table}

\section{Conclusion}
In conclusion, one needs to be careful in using wavelets for
estimating the Hurst exponent. The continuous wavelets provide an
over complete basis, because of which the coefficients are not
independent, they overestimate the power in the fluctuations. In
comparison the discrete wavelet is based on complete orthonormal
basis, ensuring that fluctuations are independent at each level.
Hence this discrete wavelet approach yields correct values for the
Hurst exponent.

{\bf Acknowledgements} We would like to thank Dr. R. Jha for
providing the tokamak plasma data for analysis.


\begin{thebibliography}{}

\bibitem{mandel} B. B. Mandelbrot, {\it The Fractal Geometry of
Nature} Freeman, San Francisco, 1999.

\bibitem{hur} H. E. Hurst, {\it Trans. Am. Soc. Civ. Eng.} {\bf 116} (1951) 770.

\bibitem{feder} J. Feder, {\it Fractals} Plenum Press, New York,
1988.

\bibitem{arn1} A. Arneodo, G. Grasseau, and M. Holshneider, {\it Phys. Rev. Lett.} {\bf 61} (1988) 2284; J. F. Muzy, E. Bacry, and A. Arneodo, {\it Phys. Rev. E} {\bf47} (1993) 875.

\bibitem{khu} K. Hu, P. Ch. Ivanov, Z. Chen, P. Carpena, and H. E. Stanley, {\it Phys.Rev. E} {\bf 64} (2001) 11114.

\bibitem{gopi} P. Gopikrishnan, V. Plerou, L. A. N. Amaral, M. Meyer, and H. E. Stanley, {\it Phys. Rev. E} {\bf 60} (1999) 5305.

\bibitem{ple} V. Plerou, P. Gopikrishnan, L. A. N. Amaral, M. Meyer, and H. E. Stanley, {\it Phys. Rev. E} {\bf 60} (1999) 6519.

\bibitem{chen} Z. Chen, P. Ch. Ivanov, K. Hu, and H. E. Stanley, {\it Phys. Rev. E} {\bf 65} (2002) 041107.

\bibitem{matia} K. Matia, Y. Ashkenazy, and H. E. Stanley, {\it Europhys. Lett.} {\bf 61} (2003) 422.

\bibitem{krs} R. C. Hwa, C.B. Yang, S. Bershadskii, J.J. Niemela, and K. R. Sreenivasan, {\it Phys. Rev. E} {\bf 72} (2005) 066308.

\bibitem{phand} K. Ohashi, L. A. N. Amaral, B. H. Natelson, and Y. Yamamoto, {\it Phys. Rev. E.} {\bf 68} (2003) 065204(R).

\bibitem{xu} L. Xu, P. Ch. Ivanov, K. Hu, Z. Chen, A. Carbone, and H.E. Stanley, {\it Phys. Rev. E} {\bf 71} (2005) 051101.

\bibitem{peng}  C. K. Peng, S. V. Buldyrev, S. Havlin, M. Simons, H. E. Stanley, and A. L. Goldberger, {\it Phys. Rev. E} {\bf 49} (1994) 1685.

\bibitem{net} J. W. Kantelhardt, D. Rybskia, S. A. Zschiegnerb,
P. Braunc, E. Koscielny-Bundea, V. Livinae, S. Havline, A. Bundea,
and H. E. Stanley, {\it Physica A} {\bf 330} (2003) 240.

\bibitem{awc}I. Simonsen, A. Hansen, and O.-M. Nes, {\it Phys. Rev. E} {\bf 58} (1998) 2779.

\bibitem{mani1} P. Manimaran, P.K. Panigrahi, and J.C. Parikh, {\it Phys. Rev. E}
{\bf72} (2005) 046120.

\bibitem{mani2} P. Manimaran, P. K. Panigrahi, and J. C. Parikh, {\it eprint: nlin.CD/0601065} (2006).

\bibitem{daub} I. Daubechies, {\it Ten lectures on wavelets} SIAM, Philadelphia, 1992.

\bibitem{mall} S. Mallat, {\it A Wavelet Tour of Signal Processing} Academic Press, 1999.

\bibitem{ram} C. S. Burrus, R. A. Gopinath, and H. Guo, {\it
Introduction to Wavelets and Wavelt Transforms} Prentise Hall,
New Jersy, 1998.

\bibitem{jos} B. K. Joseph, R. Jha, P. K. Kaw, S. K. Mattoo, C. V.
S. Rao, Y. C. Saxena, and the Aditya team, {\it Phys. Plasmas} {\bf 4} (1997) 4292.

\bibitem{jha} R. Jha, P. K. Kaw, D. R. Kulkarni, and J. C. Parikh, {\it Phys. Plasmas} {\bf10} (2003) 699.

\bibitem{rela} E. Faleiro, J. M. G\'{o}mez, R. A. Molina, L.
Mu\~{n}oz, A. Rela\~{n}o, and J. Retamosa, {\it Phys. Rev. Lett.} {\bf
93} (2004) 244101; A. Rela\~{n}o, J. Retamosa, E. Faleiro, and J.
M. G\'{o}mez, {\it Phys. Rev. E} {\bf 72} (2005) 066219.

\bibitem{ran} P. Manimaran, P. A. Lakshmi, and P. K. Panigrahi, J. Phys. A: Math. Gen. {\bf 39} (2006)
L599.
\end{thebibliography}
\end{document}